\documentstyle[11pt]{article}
\input psfig
\long\def\comment#1{}
\begin{document}
\title{Teleportation Protocols Requiring Only One Classical Bit}
\author{Subhash Kak\\
Department of Electrical \& Computer Engineering\\
Louisiana State University,
Baton Rouge, LA 70803, USA}
\maketitle

\begin{abstract}
The standard protocol for teleportation of a quantum state
requires an entangled pair of particles
and the use of two classical bits of information. Here,
we present two protocols for teleportation that require only one classical bit.
In the first protocol, chained XOR operations are performed on the
particles before one of them
is removed to the remote location where the
state is being teleported. 
In the second protocol, three entangled particles are used.
In a variant scheme, Bob's particle is distributed to him right
in the beginning (as in the standard case) and 1.5 classical
bits are required for teleportation.

\end{abstract}

\thispagestyle{empty}

\subsection*{Introduction}
In the standard teleportation protocol\cite{Be95},
an unknown quantum state  (of particle X) is teleported to a remote location
using
two entangled particles  (Y and Z) and 
two classical bits of information.
This has been interpreted as the disembodied transfer of
an unknown quantum state from one place to a remote location or
the exchange of quantum information.
(One may also consider the question of transfer of state mode
in teleportation\cite{Ka03}.)
The question of the dependence between the 
amount of classical bits of information and the conditions necessary
for teleportation to occur has not been addressed before.
Here we present new teleportation protocols that require only
one classical bit of information by altering the conditions under
which the protocols proceed.

The proposed protocols exploit the property that chained XOR
transformations correlate alternate qubits. 
In contrast to the standard
protocol where the operations are done on particles
X and Y alone, the first protocol requires operations 
involving all the three particles.
In experiments on teleportation, the entangled particles are
generated while the experiment is being performed\cite{Bo02,Id01,Ra98}, therefore, this
condition is not too restrictive.
In the second protocol, three entangled particles are used.
In a variant of the second protocol, there is no need to
transfer Bob's particle to him in the middle of the
sequence of steps.
In both these protocols, the unknown state may be
recovered back by Alice after the entangled particle has been
transmitted to the remote location.

\subsection*{The First Protocol}
Alice starts with the pure entangled state
$  | 00 \rangle +  | 11\rangle$ representing particles Y and Z
that may be assumed to have been created by a broker.
(We leave out normalizing constants in this and other expressions.)
Alice wishes to send to Bob the unknown qubit
$|\phi\rangle$ associated with X.
Without loss of generality, $|\phi\rangle = a |0\rangle + b |1\rangle$, where 
$a$ and $b$ are unknown coefficients. The initial state of the three
particles is:

\vspace{0.2in}

$ a |000\rangle + b |100\rangle + a |011\rangle + b |111\rangle $

\vspace{0.2in}

\noindent
The protocol consists of the following four steps:

\begin{enumerate}

\item Apply chained XOR transformations:

1a. XOR the states of X and Y.

1b. XOR the states  of Y and Z.

{\it The particle Z is now transferred to Bob at the remote location.}

\item Apply H on the state of X.

\item
Measure the state of X and transfer information regarding it to Bob.

\item
Apply appropriate operator on Z to complete teleportation of the
unknown state.

\end{enumerate}

\noindent
{\it Note} Steps 1b and 2 may be exchanged. However, doing so
will preclude the remote transfer of particle Z.

\paragraph{Proof}
Consider the 
\vspace{0.2in}
 $XOR =                        \left[ \begin{array}{cccc}
1 & 0 & 0 & 0 \\
0 & 1 & 0 & 0 \\
0 & 0 & 0 & 1 \\
0 & 0 & 1 & 0 \\
\end{array} \right]$
operator
on the
first two qubits (X and Y) (Step 1a). This leads to the state:

\vspace{0.2in}

$  a |000\rangle + b |110\rangle + a |011\rangle + b |101\rangle $

\vspace{0.2in}
\noindent
The next XOR operation on the
qubits Y and Z (Step 1b)  gives us the state:

\vspace{0.2in}

$  a |000\rangle + b |111\rangle + a |010\rangle + b |101\rangle $

\vspace{0.2in}
This second XOR operation makes the qubits X and Z to become fully
entangled.
This is a consequence of the property that chained XOR transformations
correlate alternate qubits.
(The power of chaining may be seen by considering the compound state
of $a |0\rangle + b |1\rangle$ and
the entangled bits 
$  | 0000 \rangle +  | 1111\rangle$ on which chained XOR
transformations (on the first and second, followed by second and
third, and so on) are applied. This gives:

\vspace{0.2in}

$ a |00000\rangle + b |11111\rangle + a |01010\rangle + b |10101\rangle $

\vspace{0.2in}
\noindent
which is characterized by symmetry and the correlations across the
first, the third, and the fifth qubits.)

The application of 
$H = \frac{1}{\sqrt 2}                       \left[ \begin{array}{cc}
1 & ~1  \\
1 & -1  \\
\end{array} \right]$
operator on the first qubit (Step 2), gives us:
\vspace{0.2in}

$ a (|000\rangle +  |100\rangle ) + b  (|011\rangle - |111\rangle) $

$ + a  (|010\rangle +  |110\rangle ) + b ( |001\rangle -  |101\rangle) $

\vspace{0.2in}
\noindent
Simplifying, we obtain:

\vspace{0.2in}

$ |00\rangle ( a |0\rangle + b |1\rangle )
+ |01\rangle (a |0\rangle + b |1\rangle )$
\vspace{0.1in}

$ + |10\rangle (a |0\rangle - b |1\rangle )
+ |11\rangle (a |0\rangle - b |1\rangle )$

\vspace{0.1in}
$ =  |0\rangle (  |0\rangle +  |1\rangle )
 (a |0\rangle + b |1\rangle )  + |1\rangle ( |0\rangle +  |1\rangle )
(a |0\rangle - b |1\rangle )$

\vspace{0.2in}

Alice now measures the first two qubits (X and Y) (Step 3).
The state of the remaining qubit (Z)
collapses to one of the two states:

\vspace{0.2in}
$a |0\rangle + b |1\rangle$ or $a |0\rangle - b |1\rangle$. 

\vspace{0.2in}
The information of the first qubit (X) is enough 
to determine which
of the two operators 
$\left[ \begin{array}{cc}
              1  & 0  \\
              0 & 1 \\
               \end{array} \right]$,
$\left[ \begin{array}{cc}
              1  & 0\\
              0 & -1 \\
               \end{array} \right]$
should be applied to Z to place it in the
state $|\phi\rangle$ (Step 4).
The measurement of the second qubit (Y) does not provide any useful
information because the qubits Y and Z are uncorrelated.

Although, the conditions for this protocol are more restrictive than
for the standard protocol, the situation with respect to Alice
and Bob is symmetric in the end. This means that if
Bob were to apply the H operator on Z instead of Alice
applying it on X, the state $|\phi\rangle$ can be recovered
by Alice (after Bob supplies the necessary one classical bit to her 
to know which transformation to apply to her state).
This supports the point of view
 that this protocol be considered a case
of true teleportation.

\subsection*{The Second Protocol}

In this protocol there are a total of four particles:
X, Y, Z, and U.
Of these, U is at the remote location 
with Bob in the closing stages of the protocol (in a variant
scheme described later, U remains with Bob right from the beginning).
The three particles Y, Z and U are in the pure
entangled state
$  | 000 \rangle +  | 111\rangle$.

The initial state of the four particles is:

\vspace{0.2in}

$ a |0000\rangle + b |1000\rangle + a |0111\rangle + b |1111\rangle $

\vspace{0.2in}

\noindent
The protocol consists of the following six steps:

\begin{enumerate}

\item Apply chained XOR transformations on the particles available to Alice:

1a. XOR the states of X and Y.

1b. XOR the states  of Y and Z.

\item Apply H on the state of X.

\item
Measure the state of X and Y.

\item
Apply appropriate operators (described in the proof) on Z and U.
U is now transferred to Bob at the remote location.

\item
Apply H operator on Z.

\item
Measure Z and transmit one classical bit of information
to Bob to complete the teleportation of $| \phi\rangle$
to him.

\end{enumerate}

\paragraph{Proof}
Consider the 
 $XOR$
operator
on the
first two qubits (X, Y) (Step 1a). This leads to the state:

\vspace{0.2in}

$  a |0000\rangle + b |1100\rangle + a |0111\rangle + b |1011\rangle $

\vspace{0.2in}
\noindent
The next XOR operation on the
qubits Y and Z (Step 1b)  gives us the state:

\vspace{0.2in}

$  a |0000\rangle + b |1110\rangle + a |0101\rangle + b |1011\rangle $

\vspace{0.2in}

The application of 
$H$
operator on the first qubit (Step 2), gives us:
\vspace{0.2in}

$ a (|0000\rangle +  |1000\rangle ) + b  (|0110\rangle - |1110\rangle) $

$ + a  (|0101\rangle +  |1101\rangle ) + b ( |0011\rangle -  |1011\rangle) $

\vspace{0.2in}
\noindent
Simplifying, we obtain:

\vspace{0.2in}

$ |00\rangle ( a |00\rangle + b |11\rangle )
+ |01\rangle (a |01\rangle + b |10\rangle )$
\vspace{0.1in}

$ + |10\rangle (a |00\rangle - b |11\rangle )
+ |11\rangle (a |01\rangle - b |10\rangle )$

\vspace{0.2in}

Alice measures the first two qubits (X and Y) (Step 3).
The state of the remaining qubits (Z and U)
collapses to one of the four states:

\vspace{0.2in}
$a |00\rangle + b |11\rangle$,
$a |01\rangle + b |10\rangle$,
$a |00\rangle - b |11\rangle$,
$a |01\rangle - b |10\rangle$.

\vspace{0.2in}
Alice uses the information in the first two
qubits to
determine which
of the four operators 
$\left[ \begin{array}{cc}
              1  & 0  \\
              0 & 1 \\
               \end{array} \right]$,
$\left[ \begin{array}{cc}
              1  & 0\\
              0 & -1 \\
               \end{array} \right]$,
$\left[ \begin{array}{cc}
              0  & 1  \\
              1 & 0 \\
               \end{array} \right]$,
$\left[ \begin{array}{cc}
              0  & 1\\
              -1 & 0 \\
               \end{array} \right]$
should be applied on Z and U so that they
are in the compound state
$a |00\rangle + b |11\rangle$.

Alice applies H on the state of Z, 
that is on $a |00\rangle + b |11\rangle$,
to give:

\vspace{0.2in}
$a|00 \rangle + a |10\rangle + b |01\rangle - b|11\rangle$

\vspace{0.1in}
$ = |0\rangle (a|0 \rangle + b |1\rangle) + |1\rangle( a|0\rangle - b|1\rangle)$.

\vspace{0.2in}

Alice measures Z, and, based on her measurement, transmits one classical
bit of information to Bob, enabling him to use the appropriate operator
on U
to obtain the unknown $|\phi\rangle$.

\subsection*{Variant of Protocol 2}
In this variant, particle U is distributed to Bob in the
beginning of the protocol. 
Therefore, this corresponds to the standard condition of
teleportation.

The protocol proceeds in exactly the same way as before until
 Step 4 where one may require a transformation
on U with probability half (when at the end of Step 3 the states
$a |01\rangle + b |10\rangle$ or
$a |01\rangle - b |10\rangle$ were obtained).
In this case a
classical bit of information is sent to Bob to
apply to his particle
$\left[ \begin{array}{cc}
              0  & 1  \\
              1 & 0 \\
               \end{array} \right]$,
the appropriate operator in this case, so that the joint state 
of Z and U is 
$a |00\rangle + b |11\rangle$.
Since this will happen only in 50\% of the cases, its
computational burden is one-half bit.
With the further requirement of one classical bit 
in Step 6, one needs a total of 1.5 classical bits.
This is better than the 2 classical bits of the standard
scheme.

\subsection*{Conclusion}

In any visualization of a teleportation experiment, it is 
convenient to generate the entangled pair at about the
same time as the teleportation is sought to be done. In such a
case, the additional pre-processing (the second XOR on Y and Z)
may not be a problem.
The first protocol works by making X and Z entangled and then
using the gate H to expand this entanglement into
the two Bell states that are teleported to Z.
In the second protocol, three entangled particles 
are required
to teleport the unknown state to the remote location.

The power of the protocols springs from their ability to transform
$a|0\rangle + b|1\rangle$ into 
$a|00\rangle + b|11\rangle$. This process may be generalized further
for several interesting effects.
The symmetry of $a|00\rangle + b|11\rangle$ from the point of view
of Alice and Bob means that the unknown state is jointly shared and
it may be recovered back by Alice.

The reduction from the four Bell states to two in the
teleported state might make it easier to implement
and verify the protocols (although the use of three
entangled particles in the second protocol would impose
additional burden).
This reduction
is evidently optimal, because otherwise
it would become possible to transfer information faster
than the speed of light.

\paragraph{Acknowledgement} I am thankful to 
Ian Glendinning and David Mermin
for their comments.

\end{document}